\documentclass[aps,twocolumn,showpacs]{revtex4}
\usepackage{amssymb}
\usepackage{amsmath}
\usepackage{epsfig}
\newcommand{\sech}{\rm sech}

\begin{document}
\title{\bf Nonlinear Landau-Zener Tunnelling in Coupled Waveguide Arrays}
\author {Ramaz Khomeriki}
\affiliation {Physics Department, Tbilisi State University, 3
Chavchavadze, 0128 Tbilisi (Georgia) \\ Max-Planck-Institut fur
Physik komplexer Systeme, N\"othnitzer str. 38, 01187 Dresden
(Germany)}
\begin{abstract}
The possibility of direct observation of Nonlinear Landau-Zener
tunnelling effect with a device consisting of two waveguide arrays
connected with a tilted reduced refractive index barrier is
discussed. Numerical simulations on this realistic setup are
interpreted via simplified double well system and different
asymmetric tunnelling scenarios were predicted just varying
injected beam intensity.
\end{abstract}
\pacs{42.65.Wi, 42.82.Et, 03.65.–w, 05.45.-a} \maketitle

Landau-Zener tunnelling effect \cite{landau} has been first
proposed for interpretation of atomic level mixing in
predissociasion process \cite{book}. More recently this model has
been applied to explain transitions between Bloch modes in
periodic systems, particularly for Bose-Einstein Condensates (BEC)
\cite{kasevich,kristiani} and acoustic waves in layered and
elastic structures \cite{kosevich}. Later on the same effect of
Bloch mode transitions has been investigated in optical systems
with variety of architectures: waveguide arrays with a step in a
refractive index \cite{ramazzener}, arrays with applied
temperature gradient \cite{trompeter}, curved waveguides
\cite{longhi}, nematic crystals \cite{assanto} and two dimensional
photonic lattices \cite{trompeter1}, among others.

The nonlinear extension of Landau-Zener model is first analyzed
theoretically \cite{wu,liu} in case of BEC in optical lattices and
asymmetric transition processes have been found in contrast to the
linear limit. Later on this nontrivial behavior has been
experimentally confirmed \cite{jona} and generalized Landau-Zener
transition formula has been analytically derived \cite{anal}. In
all of these previous studies the tunnelling processes between
different Bloch modes are considered, while, in principle, the
same Landau-Zener tunnelling effect should take place for double
well system. Indeed, very recently the tunnelling scenarios
between two spatial modes of BEC has been proposed
\cite{BECref,BECref1}. In the present paper we propose to use two
coupled waveguide arrays with a tilted (with respect to the
waveguide direction) reduced refractive index barrier for visual
observation of asymmetric nonlinear tunnelling effects between the
arrays.

The refraction index profile of a suggested experimental device is
presented in insets of Fig. 1 and refractive index pattern could
be realized either by microfabrication \cite{silberberg} or by
laser beams in photonic lattices \cite{trompeter1}. The beam is
injected either into the left or right array and its intensity has
a harmonic profile across the injection array (this is
schematically represented as black arrows with different lengths
in the insets). In case of small intensity (linear regime) the
light injected into the left array tunnels to the right array and
vice versa. Increasing the injected beam intensity the symmetry
breaks down, particularly, injecting the beam at the left it again
tunnels to the right array, while injecting the beam with the same
intensity at the right it stays trapped there (see the
corresponding graphs in Fig. 1). Further increase of the injected
intensity leads to the beam trapping irrespective to the place of
the injection of the beam. Below I shall interpret this nontrivial
effect via Nonlinear Landau-Zener tunnelling in a simple two
degree of freedom system.
\begin{figure}[t] \centerline
{\epsfig{file=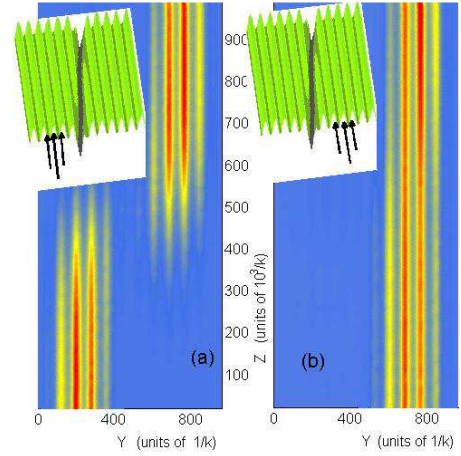,width=4cm}} \caption {(Color online) Graphs
(a) and (b): results of numerical simulations (stationary spatial
light intensity distribution) on the initial model \eqref{psi}
with the amplitude 0.0018 of injected light wave pattern
$\Psi(Y,0)$. Insets show the device schematics and corresponding
injection positions.} \label{fig:map0}\end{figure}

Let us start from writing a wave equation for linearly polarized
electric field in paraxial approximation:
\begin{equation}\label{psi}
i\frac{\partial{\Psi}}{\partial Z}+\frac{1}{2}\frac{\partial^2
{\Psi}}{\partial Y^2}+\Delta (Y,Z){\Psi}+|{\Psi}|^2{\Psi}=0,
\end{equation}
where I assume light propagation in nonmagnetic optical film ($YZ$
plane) along $Z$ direction, ${\Psi}(Y,Z)$ stands for a complex
wave envelope, $\Delta(Y,Z)=(n-n_0)/n_0$ is a linear refractive
index variation along $Y$ and $Z$ axis, the dimensionless spatial
variables $Y$ and $Z$ are scaled in units of inverse carrier
wavenumber $1/k$, and this wavenumber is defined as
$k=n_0\omega/c$ with $\omega$ being laser beam frequency and
focusing Kerr nonlinearity is scaled to unity. Let us consider
pinned boundary conditions (i.e. ${\Psi}(Y,Z)=0$ for
$Y=Y_{min},Y_{max}$), while periodic modulations together with
tilted reduced refractive index barrier (see insets of Fig. 1) are
modelled as follows:
\begin{equation}\label{tanh}
\Delta(Y,Z)=\delta n_1\sin^2(KY)-\delta n_2 \sech[(Y-\Gamma
Z)/\Lambda],
\end{equation}
$\delta n_1$ and $\delta n_2$ stand for the amplitudes of periodic
(scaled by $K$) modulation and barrier, respectively, while
$\Gamma$ and $\Lambda$ define a tilt angle and a width of the
reduced refractive index barrier. In numerical simulations on the
model \eqref{psi} the following values are fixed: $\delta
n_1=0.002$ and $\delta n_2=0.003$.

As well established, the problem of periodic array of effective
waveguides could be simplified via tight binding discretization
procedure \cite{kivshar,smerzi} when one can present the envelope
wavefunction $\Psi(y,z)$ as an expansion over approximate gaussian
eigenmodes of individual waveguides:
\begin{eqnarray}\label{tight}
\Psi(Y,Z)=\sum\limits_j{\cal E}_j(Z)\phi_j(Y) \quad \phi_j(Y)\sim
e^{-K\sqrt{\delta n_1/8}(Y-R_j)^2} \nonumber
\end{eqnarray}
where $j$ numbers waveguide center positions and thus $R_j=\pi
(2j+1)/2K$. Then one gets a Discrete Nonlinear Schr\"odinger
(DNLS) equation representation of the problem:
\begin{equation}\label{DNLS}
i\frac{\partial{\cal E}_j}{\partial Z}+\frac{C}{2}\left({\cal
E}_{j+1}+{\cal E}_{j-1}\right)-V(j,z){\cal E}_j+\chi|{\cal
E}_j|^2{\cal E}_j=0,
\end{equation}
where coupling constant $C$ is calculated from the overlapping
integrals between neighboring waveguide eigenmodes, while $\chi$
counts only the nonlinear overlap integral of the single
eigenmode. Effective potential barrier $V(j,Z)$ could be
approximated as (see Ref. \cite{ramaz}):
\begin{equation}\label{V}
V(j,Z)=V_0\sech\left[(\pi j-\Gamma K Z)/K \Lambda\right],
\end{equation}
where $V_0\sim\delta n_2$ is a potential barrier height. Further
reduction of \eqref{DNLS} is made associating $j$ with a
continuous new spatial variable $y=YK/\pi$ and then defining $
z=CZ$ and rescaling ${\cal E}\rightarrow {\cal
E}\sqrt{\chi/C}\exp[iCZ]$, $V\rightarrow V/C$ one gets Nonlinear
Scr\"odinger (NLS) equation in an external double well potential
with "moving" barrier:
\begin{equation}\label{NLS}
i\frac{\partial{\cal E}}{\partial z}+\frac{1}{2}\frac{\partial^2
{\cal E}}{\partial   y^2}-V(  y,  z){\cal E}+|{\cal E}|^2{\cal
E}=0.
\end{equation}
\begin{figure}[t] \centerline
{\epsfig{file=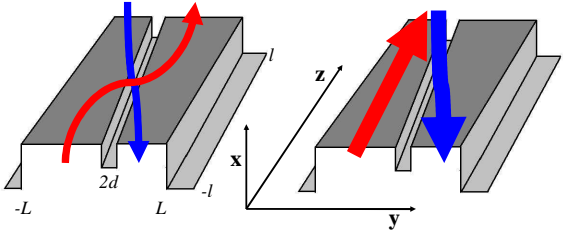,width=5cm}} \caption {(Color online) Reduction
of the initial problem \eqref{psi} with the refractive index
profiles presented in the insets of Fig. 1 to the effective double
well system with "moving" barrier. The curved lines describe the
light intensity dynamics for the linear (left graph) and
asymmetric nonlinear (right graph) regimes.}
\label{fig:map}\end{figure}

The meaning of discretization of initial equation \eqref{psi}
getting DNLS \eqref{DNLS} and then subsequent continuous
approximation to NLS equation \eqref{NLS} is that we get rid of
periodical modulation of refractive index which are present in
initial equation \eqref{psi}. Indeed, in NLS \eqref{NLS} one is
left only with "moving" barrier potential and the problem is
reduced to the two coupled waveguides case (see Fig. 2). Our aim
is to give analytical consideration of the latter problem and
present the interpretation of the numerical simulations undertaken
on the initial equation \eqref{psi}.

For the clarity of presentation let us choose symmetric boundaries
$-L<  y<L$ and $-\ell_0<  z<\ell$ and we shall find the solutions
of \eqref{NLS} in case of pinned boundary conditions ${\cal E}(-L,
z)={\cal E}(L,  z)=0$. Moreover, we will require small value for
the barrier tilt parameter $\Gamma\rightarrow 0$ and then
expanding expression for the barrier \eqref{V} over small
parameter $\Gamma z$ we get
\begin{equation}\label{bar}
V(  y,  z)=V_s(  y)+\Gamma   zV_t(  y)
\end{equation}
Thus the potential barrier expression is split into symmetric
$V_s(  y)$ and antisymmetric $V_t(  y)$ parts with respect to the
inversion transformation $  y\rightarrow -  y$:
\begin{equation}\label{barr}
V_s(  y)=\frac{V_0}{\cosh[\pi  y/K\Lambda]} \quad V_t(
y)=\frac{V_0\sinh[\pi  y/K\Lambda]}{\Lambda\cosh^2[\pi
y/K\Lambda]}
\end{equation}
and let us build the stationary solutions of \eqref{NLS} with
potential $V_{s}(  y)$ considering the second term $V_{t}( y)$ as
a perturbation. Particularly, in the zero approximation we are
left with the problem of symmetric double well potential which has
two lowest eigenvalue symmetric $\Phi^+(  y)$ and antisymmetric
$\Phi^-(  y)$ orthonormalized solutions and one can simply
construct from them two functions
\begin{equation}\label{func}
\phi_1=\left(\Phi^++\Phi^-\right)/\sqrt{2}, \qquad
\phi_2=\left(\Phi^+-\Phi^-\right) /\sqrt{2}
\end{equation}
localized at the left and right wells, respectively.

Then one can separate the variables in ${\cal E}(y,z)$
establishing a dimer model as
\begin{eqnarray}\label{e5}
{\cal E}(  y,  z)=\psi_1(  z)\phi_1(  y)+ \psi_2( z)\phi_2( y),
\quad \psi_1^2+ \psi_2^2=P_t
\end{eqnarray}
and substituting this into \eqref{NLS} where potential function is
taken in the form \eqref{bar}, multiplying on $\phi_1(y)$ and
$\phi_2(y)$, then integrating over $y$ and discarding common phase
variables we recover the nonlinear Landau-Zener model \cite{wu} in
its standard form:
\begin{eqnarray}\label{zener1}
-i\frac{\partial\psi_1}{\partial   z}= \alpha z\psi_1+v\psi_2+
r\left(|\psi_1|^2-|\psi_2|^2\right)\psi_1, \nonumber \\
-i\frac{\partial\psi_2}{\partial   z}=- \alpha   z\psi_2+
v\psi_1-r\left(|\psi_1|^2-|\psi_2|^2\right)\psi_2.
\end{eqnarray}
\begin{figure}[t] \centerline
{\epsfig{file=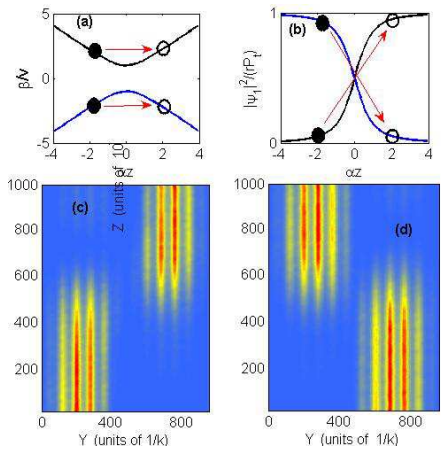,width=4cm}} \caption {(Color online) Graphs
(a) and (b) display dependencies of the propagation constant
$\beta$ and beam intensity at the left array $|\psi_1|^2$ on the
parameter $\alpha z$ according to equation \eqref{energy} in
linear regime. Solid circles indicate initial states of the system
and open circles stand for the final states in adiabatic regime.
Lower graphs display the results of numerical simulations on the
initial model equation \eqref{psi} when small intensity light is
injected into the left (graph c) and right (graph d) arrays,
respectively.} \label{fig:map3}\end{figure}
where the parameters could be calculated as follows:
\begin{eqnarray}\label{param4}
2r=\int\limits_{-L}^L d  y\phi^4_{1,2}, \quad v=
\int\limits_{-L}^L d  y\phi_{1,2}\left(\frac{\partial^2
\phi_{2,1}} {\partial   y^2}-V_{s}(  y)\phi_{2,1}\right) \nonumber
\end{eqnarray}
and $\alpha=-\Gamma\int_{-L}^L d  y \phi_{1,2} V_{t}(
y)\phi_{1,2}$ is an effective "acceleration" parameter.
\begin{figure}[b] \centerline
{\epsfig{file=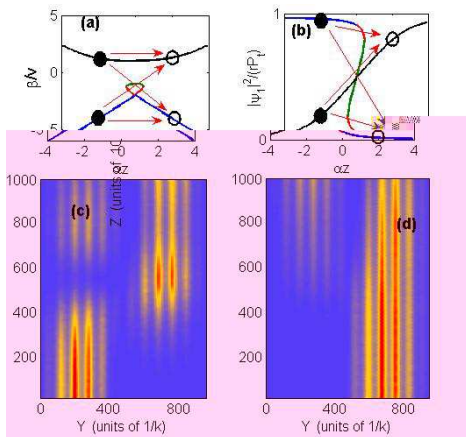,width=4cm}} \caption {(Color online) The
same as in Fig. 3 but with different parameters: Upper graphs
correspond to the intermediate range of effective nonlinearity
$rP_t=1.5$. Lower graphs display the light intensity distribution
in numerical simulations on initial equation \eqref{psi} with wave
envelope amplitude 0.0016.} \label{fig:map32}\end{figure}

In the linear limit, i.e. when total intensity $\psi_1^2+
\psi_2^2=P_t\rightarrow 0$ the above equation is just an ordinary
Landau-Zener tunnelling \cite{landau} which has a well known
result that if "acceleration" $\alpha$ is large or coupling $v$ is
small the light remains in the array where it was injected
initially. Otherwise, in adiabatic limit, the light tunnels to
other waveguide array, thus the picture is symmetric. In our
numerical simulations on the initial model equation \eqref{psi}
the barrier tilt angle $\Gamma\sim\alpha$ is small guaranteing
adiabaticity of the process and one should see tunnelling of the
light irrespective to the beam injection place.

In order to clarify the behavior of the nonlinear two degree of
freedom system \eqref{zener1} let us seek for the stationary
solutions in the form $\psi_{1,2}=|\psi_{1,2}|\exp(i\beta z)$
getting thus a quartic equation for the propagation constant
$\beta$ and dependencies of intensities $|\psi_1|^2$ and
$|\psi_2|^2$ versus $ \alpha   z$:
\begin{eqnarray}\label{energy}
( \alpha   z)^2\beta^2&=&\left(\beta^2-v^2\right)\left(\beta+rP_t\right)^2, \\
2|\psi_1|^2&=&rP_t\left(1+\frac{ \alpha z}{\beta+rP_t}\right),
\quad |\psi_2|^2=rP_t-|\psi_1|^2. \nonumber
\end{eqnarray}
From these it automatically follows that the equation for $\beta$
has four real roots if effective nonlinearity exceeds a coupling
strength $rP_t>v$, otherwise it has two real roots for fixed $z$
(for more details see Ref. \cite{wu,liu}). The corresponding
dependencies of $\beta$ and $|\psi_1|^2$ on $z$ for various total
intensities are displayed in Figs. 3, 4 and 5.
\begin{figure}[h] \centerline
{\epsfig{file=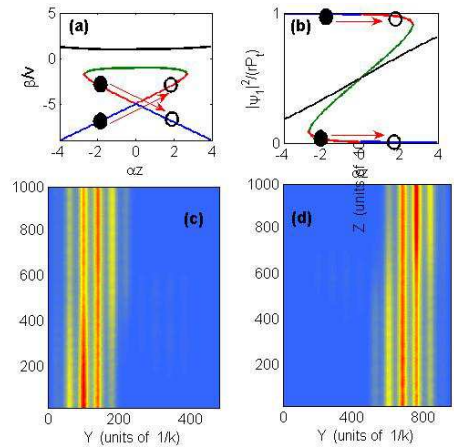,width=4cm}} \caption {(Color online) Upper
graphs correspond to the strong effective nonlinearity range
$rP_t>2$ and lower graphs display the light intensity distribution
in numerical simulations with corresponding value of wave envelope
amplitude 0.002.} \label{fig:map13}\end{figure}

Let us first consider the linear regime $rP_t\rightarrow 0$ (see
Fig. 3). Then the propagation constant is characterized by two
solutions for fixed $z$. The solid circles in upper panels of Fig.
3 correspond to the starting point $z=-\ell$ and we choose $\alpha
\ell=2$, when almost all the intensity is concentrated in the left
waveguide array. This means that we should follow blue curve and
in case of adiabatic process the system passes the point $z=0$ and
propagation constant follows further the blue curve and as a
result the system will end up with almost zero intensity at the
left waveguide array, i.e. all the intensity should go into the
right one. The same happens if the light is injected at the right
waveguide array (in this case the system is initially on the black
curve) and the light tunnels into the left array. Indeed, in
numerical simulations (see bottom panel of Fig. 3) on the initial
equation \eqref{psi} it appears that beam injection with harmonic
profile and small amplitude into the left array leads to the
tunnelling to the right array and vice versa.

Next let us consider the case when effective nonlinearity exceeds
a coupling constant $rP_t>v$. First of all we note that as seen
from the equations \eqref{energy} a simple rescaling of all
quantities with respect to the coupling constant $v$ is possible
and thus in our further analysis we can set $v=1$ without loss of
generality. Thus in case of effective nonlinearities $rP_t> 1$ the
propagation constant versus $\alpha z$ diagram acquires a
butterfly structure (see upper panels of Fig. 4) and starting
again from the point $\alpha z=-2$ and with almost whole intensity
in the left waveguide array ($|\psi_1|^2\simeq rP_t$) the
evolution along $z$ follows the blue line (see both graphs of the
upper panel of Fig. 4) then it passes to the red line region where
the butterfly structure begins [we note that different colors in
Fig. 4 corresponds to the four different solutions of quartic
equation \eqref{energy}]. By end of the red line there exists a
discontinuity, thus the system has to jump either to the blue line
or to the black one even in ideally adiabatic case. On the other
hand, starting with zero intensity at the left waveguide array
(all the intensity is concentrated at the right, i.e.
$|\psi_1|^2\simeq 0$) the system follows the black line and
continues safely until the end point $\alpha z=2$. This is the
reason of asymmetric behavior of the nonlinear case, thus
injecting the beam into different waveguide arrays one can end up
at the same array. Different amplitudes of initial harmonic beam
profiles have been checked for this tunnelling scenario. For
instance, in numerical simulations displayed in lower panels of
Fig. 4 the beam amplitude is 0.0016, while in Fig. 1 the amplitude
is 0.0018, and the asymmetric tunnelling regimes are observed.
\begin{figure}[t] \centerline
{\epsfig{file=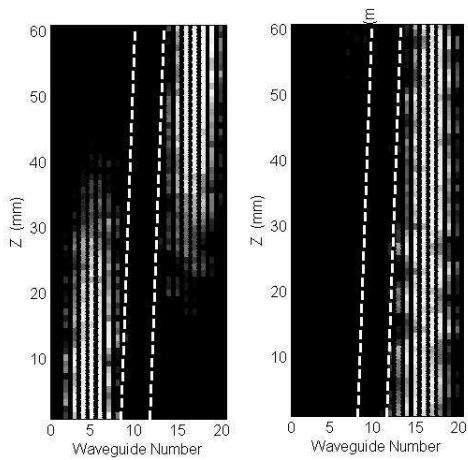,width=4cm}} \caption {Numerical simulations
on the discretized model \eqref{DNLS} with the parameters
$C=100mm^{-1}$ and $\chi=6.5mm^{-1}W^{-1}$. The and height of the
barrier and its width are $V_{max}=0.6mm^{-1}$ and $\Lambda=25\mu
m$. Dashed lines indicate position of tilted reduced refractive
index barrier and peak intensity in both cases is $|{\cal
E}_j(Z=0)|_{max}^2=31$W.} \label{fig:3223}\end{figure}

Increasing further the effective nonlinearity (in graphs of Fig. 5
we have chosen $rP_t=2$) the butterfly structure enlarges and the
symmetric behavior restores. Particularly, and this is clearly
seen from the both graphs of the bottom panel, injecting the light
in the left or right waveguide array it remain trapped there.

In order to make some predictions for realistic waveguide arrays
the most convenient way is to consider discretized equation
\eqref{DNLS}. Then one is able to unify all the variety of
refractive index profiles via a single coupling parameter $C$. In
the numerical simulations presented in Fig. 6 I choose a coupling
constant $C/2=4mm^{-1}$ and nonlinearity parameter is taken
$\chi=6.5mm^{-1}W^{-1}$ like in Ref. \cite{silberberg}. Assuming
lattice spacing equal to $6\mu m$ the tilt angle of reduced
refractive index barrier is chosen $0.25mrad$ and then the samples
of the length $60mm$ will be sufficient to see the effect of
asymmetric tunnelling. As it was mentioned above the main reason
of appearing of such asymmetry is a nonlinearity induced butterfly
structure in the reduced two degree of freedom model. In the
realistic numerical simulations on discretized equation
\eqref{DNLS} the peak intensity threshold for appearing of such a
structure is $|{\cal E}_j(Z=0)|_{max}^2=29W$ which is
experimentally easily accessible \cite{silberberg} quantity. Above
this intensity threshold the asymmetric tunnelling behavior takes
place even for fully adiabatic processes, i.e. for very small tilt
angles of reduced refractive index barrier, but then long
waveguide arrays will be required in order to see the effect.

Concluding it could be stated that visual observation of nonlinear
extension of Landau-Zener tunnelling in optical waveguide arrays
has been proposed. All analytical predictions followed from simple
two degree of freedom system are completely confirmed by numerical
simulations on the model equations \eqref{psi} and \eqref{DNLS}.

{\it Acknowledgements.} I am very grateful to D. Christodoulides
and T. Kereselidze for the useful comments. The work is supported
by Georgian National Science Foundation (Grant No GNSF/STO7/4-197)
and Science and Technology Center in Ukraine (Grant No 5053).


\begin{thebibliography}{abc}

\bibitem{landau} L. D. Landau, Phys. Z. Sowjetunion, {\bf 2}, 46 (1932);
G. Zener, Proc. R. Soc. London A, {\bf 137}, 696 (1932).
\bibitem{book} L.D. Landau, E.M. Lifshitz, {\it Quantum Mechanics, Nonrelativistic
Theory}, Moskow, Nauka (1989).
\bibitem{kasevich} B. P. Anderson and M. Kasevich, Science, {\bf 282}, 1686 (1998).
\bibitem{kristiani} M. Cristiani, O. Morsch, J. H. M\"uller, D. Ciampini, E. Arimondo,
Phys. Rev. A, {\bf 65}, 063612 (2002).
\bibitem{kosevich} H. Sanchis-Alepuz {\it et al}, Phys. Rev. Lett.
{\bf 98}, 134301 (2007); L. Gutierrez {\it et al}, Phys. Rev.
Lett. {\bf 97}, 114301 (2006);
\bibitem{ramazzener} R. Khomeriki, S. Ruffo, Phys. Rev. Lett., {\bf 94}, 113904 (2005)
\bibitem{trompeter} H. Trompeter {\it et al}, Phys. Rev. Lett., {\bf 96}, 023901 (2006)
\bibitem{longhi} F. Dreisow {\it et al}, Phys. Rev. A, {\bf 79}, 055802 (2009).
\bibitem{assanto} A. Fratalocchi, G. Assanto, Optics Express, {\bf 14}, 2021
(2006).
\bibitem{trompeter1} H. Trompeter {\it et al}, Phys. Rev. Lett., {\bf 96}, 053903 (2006)
\bibitem{wu} B. Wu and Q. Niu, Phys. Rev. A, {\bf 61}, 023402 (2000).
\bibitem{liu} J. Liu, L. Fu, B.-Y. Ou, Sh.-G. Chen, D.-I. Choi, B. Wu, Q. Niu,
Phys. Rev. A, {\bf 66}, 023404 (2002).
\bibitem{anal} D. Witthaut, E. M. Graefe, H. J. Korsch, Phys. Rev. A {\bf 73},
063609 (2006)
\bibitem{jona} M. Jona-Lasinio {\it et al}, Phys. Rev. Lett., {\bf 91}, 230406 (2003).
\bibitem{BECref} P. Engels, C. Atherton, Phys. Rev. Lett. {\bf 99}, 160405 (2007).
\bibitem{BECref1} Yu-Ao Chen {\it et al}, arXiv:1003.4956
\bibitem{silberberg} R. Morandotti  {\it et al}, Phys. Rev. Lett. {\bf 83},
4756 (1999).
\bibitem{kivshar} A.A. Sukhorukov, Y.S. Kivshar, O. Bang, C.M. Soukoulis, Phys. Rev.
E, {\bf 63}, 016615 (2000).
\bibitem{smerzi} S. Raghavan, A. Smerzi, S. Fantoni, S.R. Shenoy, Phys. Rev. A,
{\bf 59}, 620 (1999).
\bibitem{ramaz} R. Khomeriki, S. Ruffo, S. Wimberger, Europhys. Lett.,
{\bf 77}, 40005 (2007).

\end{thebibliography}
\end{document}